\documentclass{aa}
\usepackage{graphicx,natbib,amssymb}
\newcommand{\msun}{\mbox{\rm M$_{\odot}$}}
\begin{document}
   \title{New nearby stars in the Liverpool-Edinburgh high proper motion survey selected by DENIS photometry          
   \thanks{Partly based on observations made at the European Southern Observatory, La Silla, Chile}
}

   \author{C. Reyl\'e \inst{1} \and A. C. Robin \inst{1}}

   \institute{CNRS UMR6091, Observatoire de Besan\c{c}on, BP1615, 
       25010 Besan\c{c}on Cedex, France \\
   email: celine@obs-besancon.f, annie.robin@obs-besancon.fr\\
	}

   \offprints{C\'eline Reyl\'e}
   \date{Received ; accepted }

   \titlerunning{New nearby stars in the LEHPMS}

   \abstract{We present a systematic search for new neighbourhood stars among the Liverpool-Edinburgh high proper motion survey which we cross-identified with the DENIS survey. Their high proper motions ensure that they are not giant stars. The distances are estimated using DENIS photometry and we found that 100 stars probably lie within 25 parsecs from the Sun. They are mostly M-dwarfs, and 10 are probably white dwarfs. This is the first distance estimate for 84 stars among them. 10 stars, L 170-14A, DENIS J0146291-533931, DENIS J0235219-240038, LP 942-107, DENIS J0320518-635148, LTT 1732, DENIS J0428054-620929,  DENIS J2210200-701005, DENIS J2230096-534445, and NLTT 166-3 are estimated to be closer than 15 parsecs. In addition, one star, DENIS J2343155-241047, could also lie within 15~pc if it belongs to the halo.
   }

   \maketitle

\section{Introduction}
Whereas most stars more luminous than $M_V = 7$ have been detected in the solar neighbourhood \citep{jahreiss1994}, intrinsically low luminosity objects such as white dwarfs, M dwarfs and brown dwarfs may have escaped our telescopes. Yet M stars are the dominant stellar constituent of the Milky Way, and the number density of brown dwarfs may be comparable to that of stars \citep{reid1999}.

From a nearby-star inventory, \cite{henry1997} estimated that so far 130 systems are missing from the solar neighbourhood sample within 10~pc. Within 25~pc, the distance limit of the Catalogue of Nearby Stars \citep[CNS3,][]{gliese1991}, the missing fraction could be twice, that is $\sim$70\% \citep{henry2002}. These numbers were obtained on the assumption that the census is complete within 5~pc. However, recent discoveries of a M5.5 star at trigonometric distance $d = 3.7$ pc \citep{henry1997}, a M9 dwarf at spectroscopic distance $d \sim 4$ pc \citep{delfosse2001} and a M6.5 star at spectroscopic distance $d \sim 3$ pc \citep{teegarden2003} suggest that even members of the immediate solar neighbourhood remain undetected.

Projected on the sky, the apparent velocity of closer stars is greater that of farther stars. Thus, most of our known neighbours have high proper motions. From the NASA NStars database, we see that almost all of the known stars within 10~pc have a proper motion larger than 0.2$''$yr$^{-1}$ (see http://nstars.arc.nasa.gov). For many stars in high proper motion catalogues no distance has been determined yet. As part of the Research Consortium on Nearby Stars (RECONS) effort to discover new nearby stars, \cite{henry2002} obtained spectral types for high proper motion stars and candidate nearby stars and found that six of them were closer than 25~pc.

Furthermore, the new surveys in the near-infrared DENIS and 2MASS provide unprecedented data for a systematic search for low luminosity cool dwarfs and brown dwarfs. The use of these data together with existing or new high proper motion catalogues is a powerful tool for discovering our neighbours. Hundreds of stars closer than 25~pc have been discovered this way \citep{phanbao2001,phanbao2003,reid2002a,reid2002b,reyle2002,scholz2002,cruz2002,cruz2003}. Nearby stars have also been found in high proper motion catalogues from spectroscopic follow-up observations \citep{lepine2003a,rojo2003}.

In a previous paper \citep{reyle2002}, we presented the determination of the photometric distances of stars taken from the \cite{scholz2000} high proper motion catalogue and cross-identified with the DENIS survey. Near-infrared photometry was used to determine the distances. 15 new stars were found to fall within the 25~pc CNS3 limit. Spectroscopic distances should soon be obtained for selected nearby candidates from low-resolution spectroscopic observations on the New Technologies Telescope at La Silla observatory in Chile.

Recently, \cite{pokorny2003} published a large catalogue of high proper motion stars, the Liverpool-Edinburgh high proper motion survey. One of the purposes of this catalogue is to identify nearby objects. We have determined the photometric distance of the stars cross-identified with the DENIS survey.

This high proper motion catalogue is briefly described in Sect.~\ref{lehpms}. Sect.~\ref{denis} gives the result of the cross-identification with DENIS, presents the determination of photometric distances, and gives a list of the stars closer than 25~pc. The accuracy of our distance estimates is discussed in Sect.~\ref{discuss}.

\section{The Liverpool-Edinburgh high proper motion survey}
\label{lehpms}
\cite{luyten1979,luyten1980} compiled two huge high proper motion catalogues, based on observations with the Palomar Observatory Sky Survey (POSS) and the Bruce Proper Motion Survey (BPM): the Luyten Half Second proper motion catalogue (LHS) and the New Luyten Catalogue of Stars with Proper Motions Larger than Two Tenths of an Arcsecond (NLTT). Together, they contain more than 60 000 stars with proper motion higher than 0.15$''$yr$^{-1}$. Many of these stars are actually from the Lowell Observatory proper motion survey \citep{giclas1971,giclas1978}. Still, these catalogues are not complete at faint magnitudes and large proper motions \citep{dawson1986}, particularly south of declination $-33^\circ$ which is the southern limit of the POSS. Several studies have contributed to improve the completeness in the Southern hemisphere: \cite{wroblewski1997} and references therein; \cite{wroblewski1999,wroblewski2001,ruiz1993,scholz2000,ruiz2001}. In the Northern hemisphere, new large proper motion stars were also found \citep{lepine2002,lepine2003b}.

\cite{pokorny2003} published the Liverpool-Edinburgh high proper motion survey (hereafter LEHPMS), obtained using digitized plates from the ESO, UK and Palomar Schmidt surveys with the SuperCOSMOS machine. The average epoch difference between plates is 8.5 years. This catalogue contains 6206 stars over 3000 square degrees in the South Galactic Cap, with proper motions from $0.18''$yr$^{-1}$ to $20''$yr$^{-1}$ and magnitudes $9 \leq m_R \leq 19.5$, with a completeness of about 90\%. The proper motion accuracy is better than $0.1''$yr$^{-1}$ for most of the stars.

\section{Distance determination using DENIS photometry}
\label{denis}

\subsection{Cross-identification with the DENIS database}
\label{xid}
The DEep Near-Infrared Survey \citep[DENIS,][]{epchtein1997} is a survey of the Southern sky in three optical and near-infrared bands simultaneously ($I$ at 0.85$\mu$m, $J$ at 1.25$\mu$m, $K_s$ at 2.17$\mu$m). The survey, using a 1-meter telescope at ESO La Silla, started in 1996 and was completed at the end of 2001. Results for about half of the area of the Southern sky were released in May 2003. 
Source extraction and calibration have been performed at the Paris Data Analysis Center (PDAC). The astrometric accuracy is 0.5$''$ and the photometric precision is better than 0.1 mag. At the time of this work, about 80\% of the data have been processed. 

\begin{figure}[h]
\centering
\includegraphics[width=7cm,clip=,angle=-90]{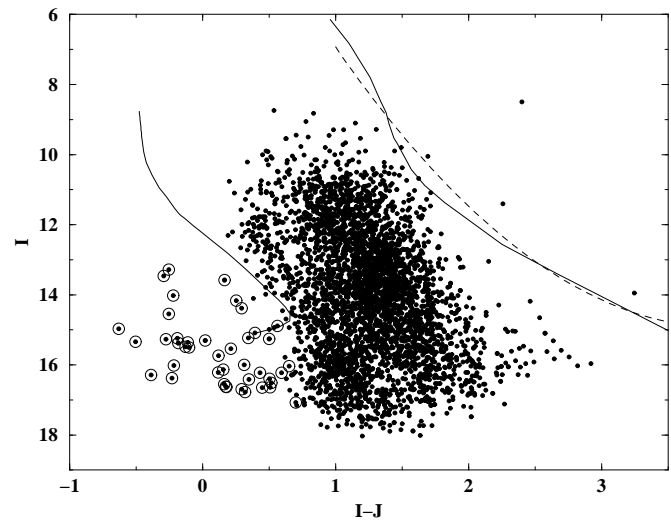}
   \caption{($I$,$I-J$) colour-magnitude diagram of the LEHPMS stars cross-identified with DENIS. Dots: G, K, or M-dwarfs. Circles: white dwarfs. Solid line with $I-J > 1$: theoretical relation for M dwarfs with solar metallicity at 10~pc \citep{baraffe1998}. Solid line with $I-J < 1$: theoretical relation for white dwarfs of 0.6~\msun~at 10~pc \citep{bergeron1995}. Dashed line: calibrated relation at 10~pc from \cite{phanbao2003}.}
   \label{fig1}
\end{figure}

We found that 4167 objects of the LEHPMS catalogue have a counterpart in the DENIS database. Most of the non-recovered stars lie in the unprocessed regions of the DENIS survey. We have checked that they have no USNO A2.0 counterpart at the position at the epoch of the DENIS observation, as expected for proper motion objects. We made the cross-identification with a search radius of 3$''$ (3$\sigma$ given the DENIS astrometric precision). Within this area, 1\% of the cross-identifications may be false, as discussed in \cite{reyle2002}. The colour-magnitude diagram of the cross-identified stars is shown in Fig.~\ref{fig1}. Three regions appear in this graph, from the lower left to the upper right: white dwarfs, halo subdwarfs, and disc main sequence stars.

\begin{figure}[h]
\centering
\includegraphics[width=7cm,clip=,angle=-90]{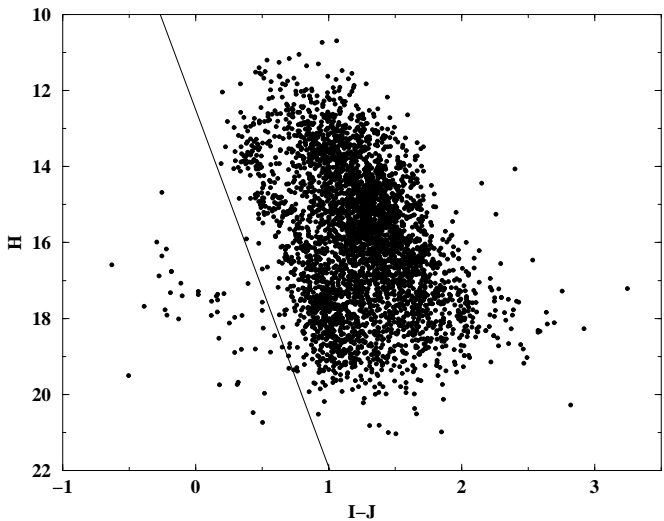}
   \caption{($H_I$,$I-J$) colour-reduced proper motion diagram of the LEHPMS stars cross-identified with DENIS.The limit we defined to separate the white dwarfs is shown by the solid line. The white dwarfs are on the left side of this line.}
   \label{fig2}
\end{figure}

According to \cite{leggett1992} all objects with $I-J > 1$ are M dwarfs or later. The reddest object ($I-J = 3.2$) is a brown dwarf. Stars with $I-J \leq 1$ are G and K dwarfs. Stars with $I-J < 0.7$ can be either distant red dwarfs or close white dwarfs. As shown by \cite{pokorny2003}, white dwarfs in the LEHPMS can be separated from red dwarfs using the reduced proper motion. Close white dwarfs are expected to have a higher proper motion at a given apparent magnitude, or to be fainter at a given distance. Thus they have higher reduced proper motions than distant red dwarfs. We isolated the white dwarfs using the reduced proper motion $H_I = I + 5 \times log\ \mu + 5$. Fig.~\ref{fig2} shows the colour-reduced proper motion diagram. Stars left of the solid line are most probably white dwarfs. Indeed, if not considered white dwarfs, their distance would be very large and consequently their tangential velocity $v_t = 4.74 \times \mu \times d$ would be above the Milky Way escape speed of about 400 km s$^{-1}$ \citep{leonard1990,meillon1997}.
Our sample contains 42 white dwarfs. However, given the photometric and astrometric uncertainties, our red dwarfs sample may still contain white dwarfs that lie on the right side of our limit and are misidentified.

\subsection{Distance determination}
Given the $I-J$ colour of each star, we computed its distance modulus and distance using ($I-J$,$M_I$) colour-magnitude relations. For the objects considered white dwarfs, we used the theoretical relation obtained from the \cite{bergeron1995} model atmosphere of 0.6\msun\ white dwarfs (solid line in the left part of the diagram in Fig.~\ref{fig1}). All white dwarfs in our sample are beyond 10~pc. For the G and K dwarfs, we used the \cite{lejeune1997} relation at solar metallicity. For the remaining objects, the M dwarfs, we used the relation at solar metallicity from \cite{baraffe1998} (solid line on the right of the diagram in Fig.~\ref{fig1}). The dashed line in Fig.~\ref{fig1} shows the relation computed by \cite{phanbao2003}, obtained by a polynomial fit on DENIS M dwarfs with known trigonometric parallaxes. The discrepancy between both relations is quite large for $1 < I-J < 2$, where the difference in absolute magnitude can be as high as 0.8~mag. The use of both relations for distance determination is discussed in Sect.~\ref{discuss}. However, the intrinsic scatter of the \cite{phanbao2003} sample is about $\pm$1 mag, and, as the authors pointed out, their sample may contain unresolved binaries that would bias their relation. We therefore chose to use the \cite{baraffe1998} theoretical relation.  

Few objects are above the solid line in Fig.~\ref{fig1}, that is closer than 10~pc. One should be aware of the saturation that occurs in the $I$ band for stars brighter than $I \simeq 9$. Thus, a wrong $I$ value of the brightest object brings it to a distance of 1.3~pc, whereas the trigonometric distance measured by HIPPARCOS is 8.8~pc \citep{perryman1997}.
The reddest object is the already known brown dwarf LP 944-20 at trigonometric distance $d = 5$~pc \citep{tinney1996}. 

97 stars have a photometric distance smaller than 25~pc. Their characteristics are given in Table~1. Approximate spectral types derived from the \cite{leggett1992} colour-spectral type relation are indicated. These stars are mostly M-dwarfs, except 10 probable white dwarfs. M-dwarfs are mainly from spectral type M3 to M6. Three objects that have a M7 spectral type are ultra-cool dwarfs, as defined by \cite{kirkpatrick1997}. Six additional ultra-cool dwarfs with spectral type M7 to M8 are found to be within 25 to 32~pc of the Sun. Although they are just beyond the CNS3 limit, we think they are important enough to be listed in Table~2.

\begin{table*}
\label{tab1}
\caption{Nearby stars candidates within 25~pc. The distance $d$ is estimated from the $I-J$ colour index and the \cite{baraffe1998} theoretical colour-magnitude relation. Indicative spectral types for M-stars are derived from the \cite{leggett1992} colour-spectral type relation. The white dwarfs (wd) candidates and possible halo subdwarfs candidates (sd) are given at the end of the table.}
\begin{tabular}{lccccccccccc}
\hline
\hline
name &$\alpha_{J2000}$ &$\delta_{J2000}$ &ESO&$\mu$ &$B_J$ &$R_{ESO}$ &$R_{UK}$ &$I$ &($I-J$) &$d$ &type\\
     &                 &                 &epoch      &(''yr$^{-1}$)&&&&\multicolumn{2}{c}{DENIS}&(pc)&\\
\hline
LTT 44 		&00 07 05.03& -56 05 04.46& 1989.60& 0.36& 14.22& 12.18& 12.03& 10.93& 1.38& 24.6& M3\\
G 267-034	&00 07 54.39& -29 58 41.74& 1981.75& 0.21& 14.11& 11.68& 11.85& 10.86& 1.37& 24.9& M3\\
		&00 09 43.08& -41 17 36.54& 1988.85& 0.23& 17.24& 14.71& 15.01& 12.49& 1.61& 24.0& M4.5\\
LTT 83 		&00 11 49.42& -55 21 48.45& 1989.60& 0.34& 13.99& 11.86& 11.93& 10.75& 1.36& 24.6& M3\\
LHS 1070 	&00 24 44.29& -27 08 31.31& 1988.84& 0.59& 17.03& 14.20& 14.42& 11.40& 2.26&  5.9& M6\\
		&00 28 54.55& -27 33 34.28& 1984.80& 0.23& 16.65& 14.63& 14.56& 12.30& 1.67& 19.9& M4.5\\
L 170-14A 	&00 29 50.33& -54 41 32.55& 1988.68& 0.35& 14.51& 12.89& 13.03& 11.50& 1.64& 14.6& M4.5\\ 
L 291-115 	&00 33 13.10& -47 33 18.56& 1984.87& 0.31& 16.96& 14.47& 14.62& 12.26& 1.79& 15.7& M5\\
LHS 1096 	&00 33 21.91& -24 25 11.66& 1988.84& 0.75& 17.53& 15.19& 15.41& 12.97& 1.78& 22.3& M5\\ 
WT 1115 	&00 37 35.70& -27 08 29.07& 1987.86& 0.25& 15.56& 13.10& 13.50& 11.49& 1.44& 24.2& M3.5\\ 
		&00 54 49.71& -18 23 17.19& 1987.75& 0.19& 15.40& 13.05& 13.21& 12.04& 1.56& 22.0& M4\\
LHS 129 	&00 58 26.84& -27 51 21.87& 1988.86& 1.34& 13.39& 11.07& 11.69&  9.28& 1.31& 16.0& M3\\
		&01 01 57.31& -55 23 47.06& 1981.80& 0.29& 15.34& 13.20& 13.33& 11.58& 1.54& 18.5& M4\\
LHS 134 	&01 04 52.69& -18 07 34.46& 1987.75& 1.39& 15.51& 12.90& 13.33& 11.10& 1.72& 10.4& M4.5\\
G 268-116 	&01 06 27.08& -26 01 22.74& 1987.89& 0.23& 16.09& 13.94& 14.03& 12.52& 1.66& 22.1& M4.5\\
LHS 1197 	&01 08 17.56& -28 48 19.02& 1986.82& 0.77& 14.55& 12.17& 12.18& 10.68& 1.62& 10.2& M4.5\\
LHS 1201 	&01 08 46.93& -37 10 16.84& 1989.75& 0.71& 17.29& 14.92& 15.24& 12.93& 1.92& 17.9& M5.5\\
		&01 10 50.24& -46 00 14.39& 1989.74& 0.20& 15.36& 13.01& 13.29& 11.85& 1.52& 22.1& M4\\
LTT 692 	&01 14 33.99& -53 56 37.30& 1981.80& 0.33& 12.21&  9.76&  9.95&  9.10& 1.15& 24.6& M1\\
NLTT 938-149 	&01 15 29.85& -33 25 49.97& 1989.75& 0.23& 17.14& 14.89& 14.99& 12.62& 1.67& 23.0& M4.5\\
		&01 20 15.55& -50 08 50.88& 1981.82& 0.25& 15.40& 13.11& 13.41& 11.83& 1.52& 21.9& M4\\
NLTT 939-44 	&01 24 30.48& -33 55 00.31& 1989.75& 0.23& 16.83& 14.48& 14.73& 12.27& 1.57& 23.4& M4\\
		&01 26 42.61& -54 43 29.39& 1984.89& 0.29& 17.27& 14.86& 15.10& 12.52& 1.62& 24.1& M4.5\\
		&01 27 31.69& -31 40 04.51& 1985.92& 0.32& 21.46& 18.75& 19.04& 15.31& 2.64& 24.8& M7\\
LP 768-25 	&01 30 43.01& -19 29 12.21& 1990.70& 0.20& 15.81& 13.38& 13.62& 12.21& 1.58& 22.5& M4\\
		&01 35 21.57& -22 14 33.52& 1991.00& 0.20& 16.69& 14.39& 14.49& 12.48& 1.70& 20.4& M4.5\\
LHS 1293 	&01 45 11.07& -32 05 09.66& 1985.93& 0.70& 15.37& 13.25& 13.22& 11.36& 1.51& 18.0& M4\\
		&01 45 14.59& -31 36 27.17& 1985.93& 0.22& 16.22& 13.77& 13.97& 12.20& 1.54& 24.5& M4\\
		&01 45 30.37& -52 30 20.11& 1984.89& 0.18& 15.26& 13.86& 14.27& 12.20& 1.71& 17.5& M4.5\\
NLTT 940-4 	&01 45 35.17& -35 10 25.04& 1985.88& 0.27& 15.22& 12.92& 12.99& 11.77& 1.52& 21.1& M4\\
		&01 46 29.07& -53 39 30.88& 1984.89& 0.21& 15.52& 13.18& 13.31& 11.03& 1.59& 12.7& M4\\
		&01 47 59.25& -27 42 16.12& 1985.93& 0.22& 16.70& 14.28& 14.51& 12.61& 1.62& 24.8& M4.5\\
LP 940-20 	&01 49 42.26& -33 19 21.37& 1985.88& 0.44& 17.00& 14.87& 14.85& 12.65& 1.77& 19.7& M5\\
		&02 07 13.59& -37 21 52.15& 1986.81& 0.44& 20.80& 18.02& 18.22& 15.10& 2.58& 23.9& M7\\
LHS 1367 	&02 15 07.38& -30 39 57.17& 1988.77& 0.84& 20.23& 17.34& 17.68& 14.18& 2.46& 17.4& M6.5\\
NLTT 829-41 	&02 16 21.55& -22 00 52.03& 1984.83& 0.25& 18.14& 15.55& 15.72& 13.21& 1.96& 19.6& M5.5\\
NLTT 941-57 	&02 22 18.07& -36 51 52.80& 1989.81& 0.31& 17.45& 15.16& 15.20& 12.98& 1.93& 18.2& M5.5\\
		&02 22 47.82& -27 32 33.70& 1988.77& 0.23& 20.65& 17.95& 18.23& 14.65& 2.53& 20.2& M7\\
LP 829-51 	&02 23 24.24& -23 18 20.80& 1985.69& 0.35& 15.34& 13.30& 13.14& 11.35& 1.43& 24.1& M3\\
NLTT 993-43 	&02 27 58.74& -42 43 33.45& 1978.89& 0.19& 16.86& 14.44& 14.59& 13.10& 1.77& 24.0& M5\\
NLTT 941-90 	&02 28 07.38& -36 28 20.14& 1989.81& 0.48& 14.88& 12.02& 12.77& 11.82& 1.49& 23.6& M4\\
L 225-57 	&02 34 20.75& -53 05 31.51& 1983.92& 0.43& 13.57& 11.00& 11.09&  9.79& 1.49&  9.3& M4\\
		&02 38 48.96& -39 16 37.38& 1988.69& 0.19& 16.47& 14.13& 14.21& 12.18& 1.58& 22.0& M4\\
LP 993-98 	&02 42 05.71& -41 24 34.49& 1988.69& 0.42& 18.19& 15.59& 15.85& 13.22& 1.94& 20.1& M5.5\\
LP 831-1 	&02 54 39.13& -22 15 57.33& 1985.85& 0.42& 13.68& 11.34& 11.57& 10.34& 1.37& 19.9& M3\\
		&03 03 00.38& -55 24 55.31& 1983.92& 0.23& 16.29& 13.97& 14.07& 12.04& 1.52& 24.1& M4\\
LP 942-107 	&03 05 10.79& -34 05 23.15& 1983.76& 0.37& 14.41& 11.97& 11.92& 11.14& 1.55& 14.8& M4\\
LHS 1500 	&03 08 24.66& -38 12 45.06& 1981.91& 0.51& 17.71& 15.30& 15.43& 12.99& 1.83& 20.8& M5\\
		&03 08 59.98& -49 24 53.50& 1981.83& 0.19& 16.19& 13.67& 13.92& 12.13& 1.69& 17.5& M4.5\\
NLTT 994-114 	&03 09 21.59& -39 11 02.84& 1981.91& 0.38& 13.56& 11.25& 11.96& 10.36& 1.42& 15.9& M3\\
NLTT 772-8 	&03 09 50.86& -19 06 46.87& 1986.90& 0.42& 15.41& 12.85& 13.17& 11.52& 1.56& 17.4& M4\\
LP 831-45 	&03 14 17.81& -23 09 32.36& 1984.96& 0.42& 13.29& 11.05& 11.09&  9.90& 1.44& 11.7& M3\\
LHS 1524 	&03 17 17.57& -19 40 14.70& 1986.90& 0.59& 16.57& 14.23& 14.41& 12.65& 1.78& 19.3& M5\\
		&03 20 51.84& -63 51 47.87& 1984.74& 0.32& 14.29& 11.77& 11.95& 10.59& 1.48& 14.0& M3\\
		&03 20 58.44& -55 20 18.82& 1987.80& 0.39& 19.39& 17.02& 17.24& 14.30& 2.23& 23.0& M6\\
		&03 32 17.21& -54 09 36.83& 1987.80& 0.23& 17.96& 15.46& 15.80& 13.25& 1.81& 24.3& M5\\
\hline\\
\end{tabular}
\end{table*}

\begin{table*}[t]
{\bf \small Table 1.} Continued.\vspace*{0.15cm}\\
\begin{tabular}{lccccccccccc}
\hline
\hline
name &$\alpha_{J2000}$ &$\delta_{J2000}$ &ESO &$\mu$ &$B_J$ &$R_{ESO}$ &$R_{UK}$ &$I$ &($I-J$) &d &type\\
     &                 &                 &epoch     &(''yr$^{-1}$)&&&&\multicolumn{2}{c}{DENIS}&(pc)&\\
\hline
LTT 1732 	&03 38 55.65& -52 34 14.98& 1981.80& 0.25& 15.74& 13.18& 13.42& 11.26& 1.58& 14.6& M4\\
LP 944-20 	&03 39 34.93& -35 25 46.93& 1986.95& 0.45& 20.24& 16.70& 17.02& 13.95& 3.25&  7.8& M9\\
		&04 07 30.05& -40 05 29.55& 1984.84& 0.31& 17.56& 15.37& 15.42& 13.09& 1.77& 23.8& M5\\
		&04 28 05.44& -62 09 28.60& 1989.81& 0.38& 14.15& 11.76& 12.06& 10.37& 1.50& 11.9& M4\\
		&05 03 24.90& -53 53 20.93& 1982.04& 0.49& 15.61& 13.14& 13.25& 11.82& 1.64& 16.6& M4.5\\
NLTT 83-11 	&22 10 10.48& -70 10 06.84& 1984.65& 0.28& 14.03& 11.78& 11.92& 11.00& 1.46& 17.9& M3\\
		&22 10 19.97& -70 10 04.41& 1984.65& 0.27& 14.88& 12.66& 12.79& 11.56& 1.67& 14.2& M4.5\\
		&22 13 50.19& -63 42 12.36& 1984.67& 0.19& 18.28& 15.61& 15.79& 13.04& 2.15& 14.3& M6\\
LHS 3793 	&22 19 22.77& -28 23 17.05& 1985.69& 0.90& 15.79& 13.52& 13.64& 11.77& 1.62& 16.9& M4.5\\
NLTT 931-48 	&22 21 53.07& -31 31 21.24& 1985.69& 0.29& 14.56& 12.31& 12.46& 10.88& 1.41& 20.5& M3\\
LHS 3799 	&22 23 06.66& -17 36 15.35& 1984.78& 0.79& 14.23& 11.67& 11.76& 10.05& 1.69&  6.7& M4.5\\
		&22 30 09.59& -53 44 44.79& 1985.68& 0.75& 15.81& 13.39& 13.57& 11.28& 1.68& 12.0& M4.5\\
		&22 34 04.12& -61 07 40.41& 1985.53& 0.26& 16.36& 13.84& 13.68& 11.75& 1.65& 15.9& M4.5\\ 
NLTT 1033-31	&22 35 04.52& -42 17 45.30& 1981.80& 0.31& 14.40& 11.80& 12.30& 10.54& 1.40& 18.3& M3\\
LHS 3842 	&22 40 57.46& -45 43 19.50& 1983.82& 0.47& 14.86& 12.59& 12.82& 11.30& 1.55& 16.0& M4\\
NLTT 876-51 	&22 41 07.65& -22 21 53.97& 1984.81& 0.27& 15.47& 13.65& 13.47& 11.87& 1.54& 21.3& M4\\
NLTT 166-3 	&22 41 59.52& -59 15 12.25& 1986.82& 0.34& 15.34& 13.07& 13.44& 11.46& 1.71& 12.5& M4.5\\
		&22 42 00.71& -37 17 33.94& 1981.80& 0.25& 16.10& 13.56& 13.90& 11.92& 1.56& 20.5& M4\\
NLTT 877-72  	&23 09 59.91& -21 11 43.06& 1984.81& 0.34& 13.67& 10.89& 11.52& 10.22& 1.36& 19.1& M3\\
LHS 3910 	&23 12 14.61& -17 45 40.07& 1984.81& 0.53& 16.32& 14.27& 13.99& 12.21& 1.55& 24.4& M4\\
		&23 25 24.54& -67 40 05.63& 1984.82& 0.29& 15.60& 13.22& 13.32& 11.66& 1.61& 16.5& M4.5\\
LHS 547 	&23 36 51.29& -36 28 52.07& 1989.08& 1.13& 14.85& 12.15& 12.76& 10.73& 1.53& 12.8& M4\\
LP 878-73 	&23 40 47.99& -22 25 27.95& 1986.58& 0.44& 16.36& 14.10& 14.27& 12.62& 1.77& 19.2& M5\\ 
LHS 3999 	&23 41 00.02& -44 57 16.64& 1989.51& 0.77& 14.37& 11.92& 12.30& 11.03& 1.41& 22.6& M3\\
G 275-082 	&23 41 16.02& -26 57 21.86& 1986.81& 0.34& 14.22& 11.80& 12.35& 10.49& 1.37& 21.2& M3\\
NLTT 1035-50 	&23 41 28.76& -38 27 52.44& 1985.77& 0.47& 17.89& 15.62& 15.60& 13.19& 1.77& 24.8& M5\\
		&23 44 10.94& -68 31 41.80& 1984.82& 0.33& 15.67& 13.49& 13.52& 11.59& 1.50& 20.7& M4\\
NLTT 1035-64 	&23 47 56.18& -39 11 09.65& 1985.77& 0.35& 15.73& 13.45& 13.61& 11.68& 1.46& 24.3& M3\\
LTT 9783 	&23 53 08.12& -42 32 04.55& 1989.81& 0.21& 13.93& 11.24& 11.94& 10.56& 1.44& 15.9& M3\\
NLTT 987-47	&23 53 40.74& -35 59 07.56& 1988.79& 0.34& 14.77& 12.46& 12.69& 10.82& 1.43& 19.1& M3\\
		&23 58 49.37& -54 00 13.19& 1989.58& 0.22& 15.64& 13.05& 13.10& 11.79& 1.56& 19.6& M4\\
\hline
\hline
LHS 1008	&00 02 39.96& -34 13 30.98& 1988.79& 0.80& 15.65& 15.00& 15.02& 14.38& 0.29& 17.5& wd\\
		&00 09 39.52& -28 48 44.02& 1981.75& 0.35& 19.05& 17.02& 17.08& 16.03& 0.65& 20.3& wd\\
LHS 1126	&00 41 26.43& -22 20 57.48& 1987.86& 0.57& 14.56& 13.43& 13.51& 13.58& 0.16& 14.7& wd\\
		&01 24 03.54& -42 40 29.96& 1983.82& 0.62& 15.62& 14.81& 14.68& 14.16& 0.25& 16.9& wd\\
  		&01 34 52.67& -48 23 47.64& 1981.82& 0.25& 17.82& 15.99& 16.06& 15.09& 0.39& 20.6& wd\\
		&02 07 02.03& -30 23 30.61& 1985.93& 0.28& 17.55& 16.83& 16.68& 16.22& 0.59& 25.0& wd\\
		&02 18 30.93& -39 36 34.10& 1986.81& 0.52& 15.93& 15.41& 15.49& 15.23& 0.34& 23.9& wd\\
   		&02 35 21.91& -24 00 38.34& 1985.85& 0.63& 16.58& 15.36& 15.51& 14.32& 0.56& 14.3& wd\\
		&03 12 25.36& -64 44 10.16& 1986.81& 0.19& 13.10& 13.23& 13.11& 13.28&-0.25& 24.9& wd\\
 		&22 48 08.69& -47 14 45.06& 1983.82& 0.32& 18.16& 16.19& 16.38& 15.26& 0.50& 18.8& wd\\
\hline
\hline
LHS 1143& 00 45 54.79& -42 47 39.10& 1989.74& 0.81& 17.00& 14.74& 14.95& 13.35& 1.18& 23.2& sd\\
& 22 09 46.31& -56 21 28.22& 1984.56& 0.66& 15.83& 13.80& 13.84& 13.43& 1.19& 22.9& sd\\
& 23 43 15.49& -24 10 47.16& 1986.81& 2.58& 14.03& 11.72& 12.05& 11.17& 1.09& 12.3& sd\\

\hline\\
\end{tabular}
\end{table*}

\begin{table*}[t]
\label{tab2}
\caption{Ultra-cool dwarfs candidates.}
\begin{tabular}{lccccccccccc}
\hline
\hline
name &$\alpha_{J2000}$ &$\delta_{J2000}$ &ESO &$\mu$ &$B_J$ &$R_{ESO}$ &$R_{UK}$ &$I$ &($I-J$) &d &type\\
     &                 &                 &epoch     &(''yr$^{-1}$)&&&&\multicolumn{2}{c}{DENIS}&(pc)&\\
\hline
		&00 05 47.68& -21 57 17.62& 1984.79& 0.71& 22.58& 19.32& 19.42& 16.02& 2.82& 29.3& M8\\
CS2 1601 	&00 07 07.76& -24 58 03.67& 1990.73& 0.20& 21.53& 18.78& 18.56& 15.77& 2.76& 27.7& M8\\ 
		&00 51 22.92& -22 51 29.88& 1987.86& 0.35& 21.41& 18.55& 18.46& 15.62& 2.59& 30.0& M7\\
		&00 58 06.91& -53 18 04.76& 1981.80& 0.32& 21.16& 18.47& 18.57& 15.59& 2.70& 26.7& M7.5\\
		&22 48 50.55& -26 41 59.74& 1984.81& 0.29& 21.41& 18.54& 18.60& 15.83& 2.64& 31.3& M7\\
		&23 56 10.76& -34 26 01.19& 1988.79& 0.29& 22.21& 19.02& 19.31& 15.96& 2.92& 26.2& M8\\
\hline\\
\end{tabular}
\end{table*}

\section{Discussion}
\label{discuss}
\begin{figure}[h]
\centering
\includegraphics[width=7cm,clip=,angle=-90]{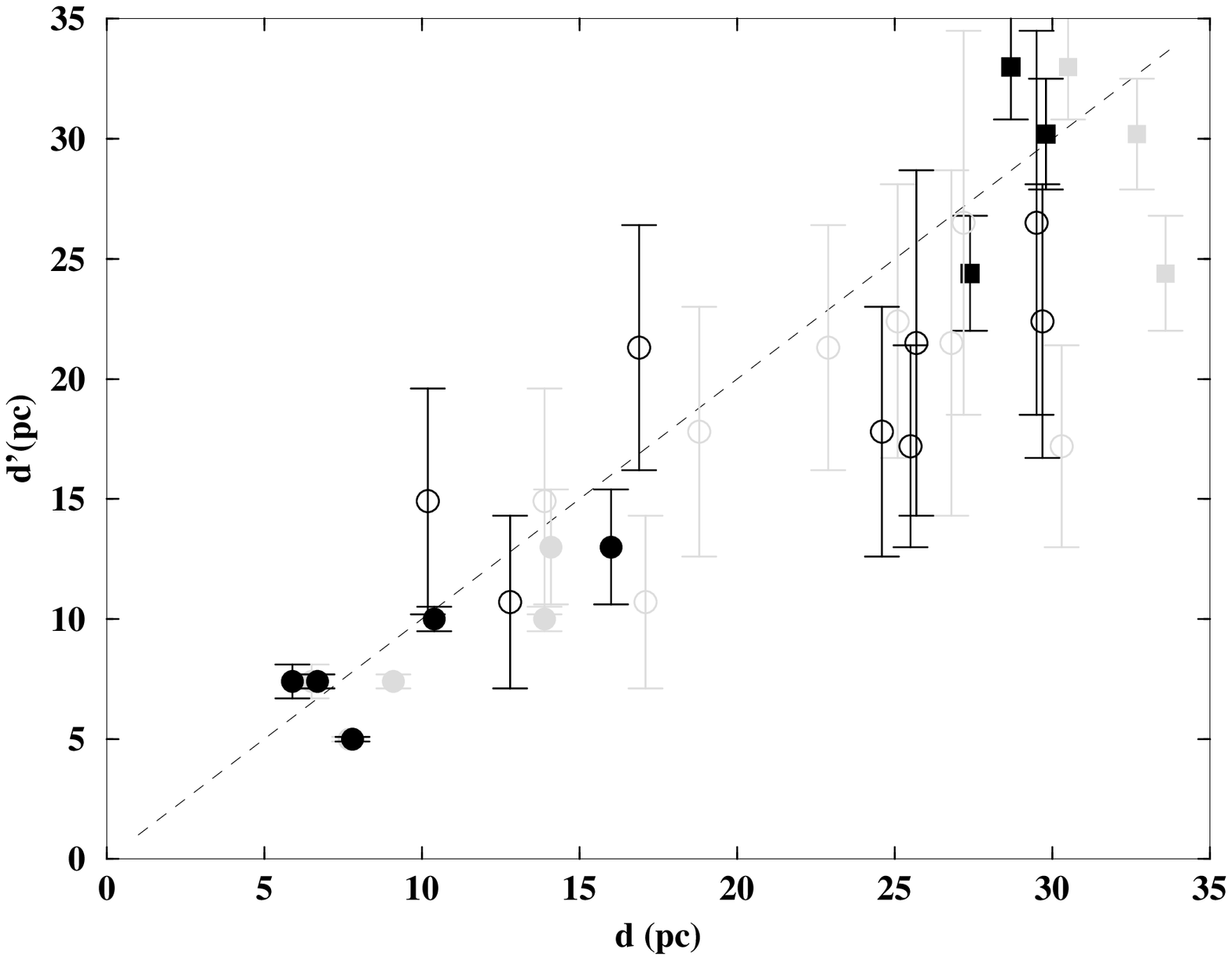}
\includegraphics[width=7cm,clip=,angle=-90]{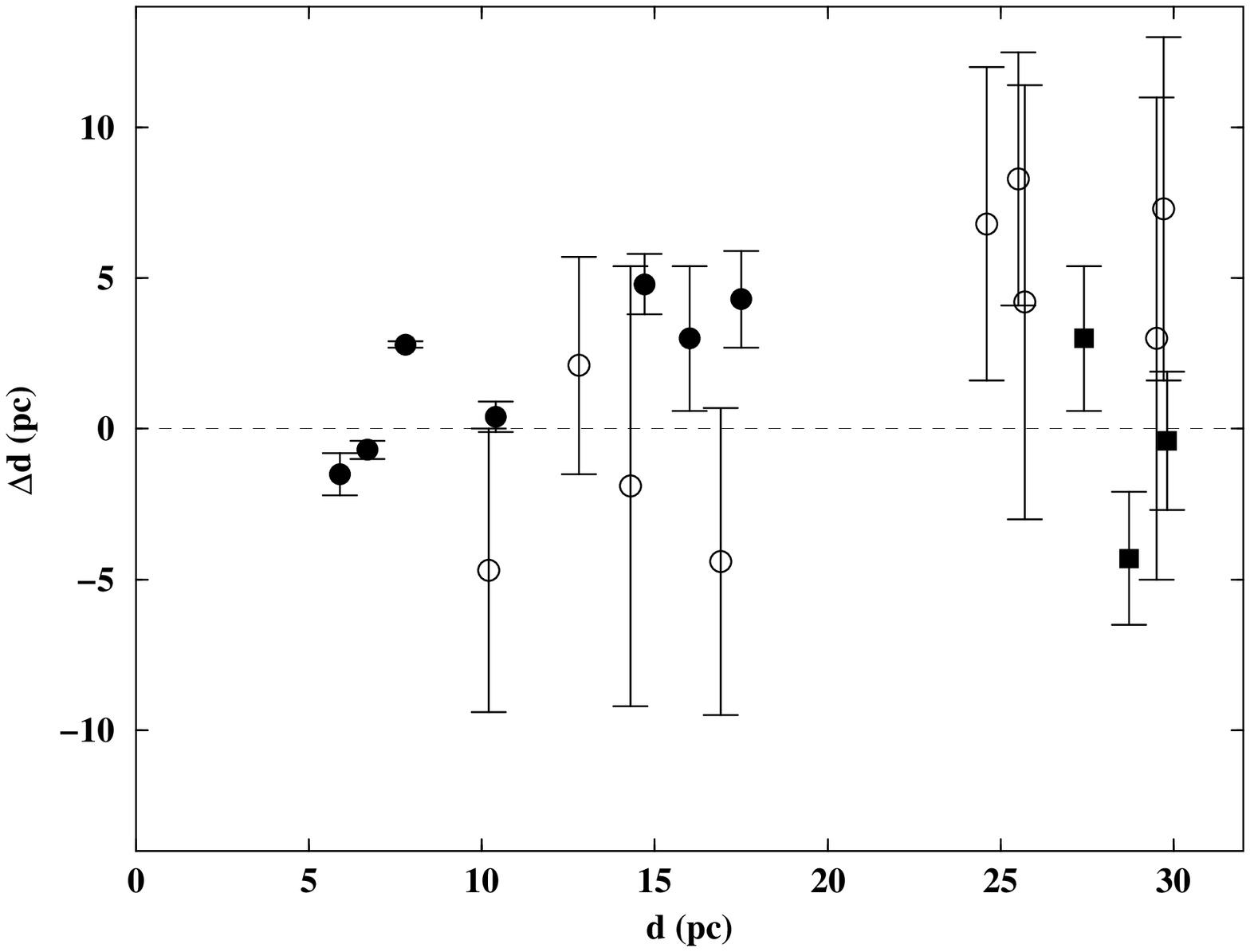}
   \caption{Upper panel: already published determination $d'$ versus our determination $d$.  Black symbols: $d$ obtained with the \cite{baraffe1998} relation. Gray symbols: $d$ obtained with the \cite{phanbao2003} relation. Filled circles: trigonometric determination of $d'$. Open circles: photometric determination of $d'$. Filled squares: spectroscopic determination of $d'$. Lower panel: difference $\Delta d$ between our distance determinations and other determinations as a function of our determination $d$.}
   \label{fig3}
\end{figure}

Our results include stars with a previously known distance, 17 of them being among our candidates closer than 25~pc. Their distance and the method of its determination are given in Table~3, as well as our determinations using \cite{baraffe1998} and \cite{phanbao2003} ($I-J$,$M_I$) colour-magnitude relations. The upper panel in Fig.~\ref{fig3} shows already published values $d'$ versus our value $d$. The different symbols indicate different methods of determination: trigonometric (filled circles), photometric (open circles) or spectroscopic (filled squares). The error bars are those of the determination from other authors. Black symbols indicate when the \cite{baraffe1998} relation is used, and gray symbols when the \cite{phanbao2003} relation is used. Neither of the two relations seems to give results that are significantly better according to already published distances, in particular trigonometric distances which are the most accurate. Indeed, the standard deviation of the difference between our distances and other determinations is 4 pc regardless of the colour-magnitude relation we used. The mean of the difference is larger when we use the \cite{phanbao2003} relation (3 pc) than when we use the \cite{baraffe1998} relation (1.5 pc). This could be due to the presence of unresolved binaries in the \cite{phanbao2003} sample.

The comparison between published distances and our determination obtained with the \cite{baraffe1998} relation is shown in the lower panel of Fig.~\ref{fig3}. The difference $\Delta d$ is plotted as a function of our distance $d$. The stars which distance has been determined by \cite{reyle2002} are not included in the plot as the values are identical, the method used being the same. The meaning of symbols and error bars is the same as before. The plot tends to show that our method suffers no systematic bias. The largest differences are with other photometric distances for which the errors are large and that may be biased.
The errors in our estimations are also large. An error of 1 mag in the absolute magnitude $M_I$ can lead to an error in the distance as large as 45\%. On the one hand, given the uncertainties on the DENIS colour, the error on the absolute magnitude can reach 1 mag in the steepest part of the colour-magnitude relation, for $I-J \leq 1.5$, that is for M4 stars and earlier. For stars later than M4, the error is twice as small. On the other hand, the uncertainties in the ($I-J$,$M_I$) relations can also be of the order of 1 mag.

\begin{table*}
\caption{Stars with previously known distance $d'$. The method used to determine the distance is indicated. Our two determinations obtained with \cite{baraffe1998} and \cite{phanbao2003} colour-magnitude relations are also given for comparison.}
\begin{center}
\begin{tabular}{lcrrcc}
\hline
\hline
name		&\multicolumn{3}{c}{other determination}  &\multicolumn{2}{c}{our determination $d$ (pc) using}	\\
		&$d'$ (pc)  &method		&ref.&Baraffe et al. &Phan Bao et al. \\
\hline
LHS 1008	&13.2	&trigonometric		&1   &17.5$^*$ &---  \\   
LHS 1070	&7.4	&trigonometric		&1   &5.9  &6.5  \\
L 291-115	&15.7   	&photometric	&2   &15.7 &20.8 \\
LHS 1126	&9.9	&trigonometric		&1   &14.7$^*$ &---  \\
LHS 1146	&30.2	&spectroscopic		&3   &29.8 &32.7 \\
LHS 129		&13.0	&trigonometric		&1,4 &16.0 &14.1 \\
LHS 134		&10.0	&trigonometric		&1   &10.4 &13.9 \\
LHS 1197	&14.9	&photometric		&5   &10.2 &13.9 \\
LP 826-603	&17.2	&photometric		&5   &25.5 &30.3 \\
LHS 1217	&21.5	&photometric		&5   &25.7 &26.8 \\
LTT 692		&17.8	&photometric		&5   &24.6 &18.8 \\
LP 940-20	&19.6	&photometric		&2   &19.6 &25.8 \\
L 223-77	&29.6	&photometric		&2   &29.6 &35.8 \\
LHS 1408	&22.4	&photometric		&5   &29.7 &25.1 \\
L 225-57	&9.6	&photometric		&2   &9.6  &12.0 \\
LP 831-1	&21.3	&photometric		&2   &21.3 &20.0 \\
LP 837-37	&33.0	&spectroscopic		&3   &28.7 &30.5 \\
LP 831-45	&11.7	&photometric		&2   &11.7 &14.1 \\
LP 944-20	&5.0	&trigonometric		&6   &7.8  &7.7  \\
J2213504-634210	&16.2	&photometric		&7   &14.3 &16.2 \\
LHS 3793	&21.3	&photometric		&5   &16.9 &22.9 \\
LHS 3799	&7.4	&trigonometric		&1   &6.7  &9.1  \\
LHS 3970	&24.4	&spectroscopic		&3   &27.4 &33.6 \\
LHS 547		&10.7	&trigonometric		&1   &12.8 &17.1 \\
LHS 4038	&26.5	&trigonometric		&1   &29.5 &27.2 \\
LP 987-60	&28.9	&photometric		&2   &28.9 &31.2 \\
\hline
\multicolumn{6}{l}{* distance obtained with the \cite{bergeron1995} colour-magnitude relation for white dwarfs.}\\
\end{tabular}
\end{center}
1- Yale Parallax Catalogue, 2- \cite{reyle2002}, 3- \cite{cruz2002}, 4- HIPPARCOS catalogue, 5- CNS3, 6- \cite{tinney1996}, 7- \cite{phanbao2001}
\end{table*}

\begin{figure}[h]
\centering
\includegraphics[width=7cm,clip=,angle=-90]{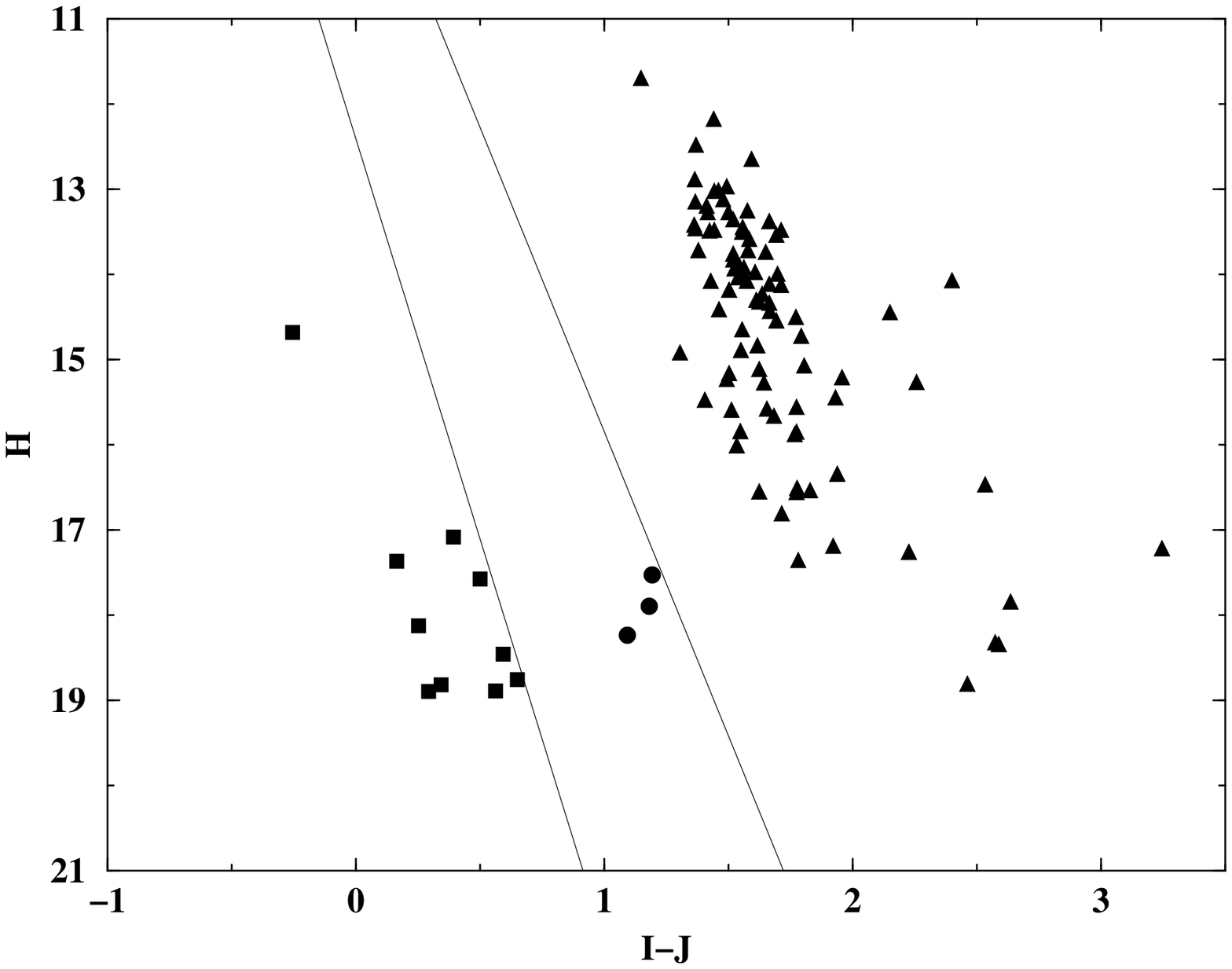}
   \caption{($H_I$,$I-J$) colour-reduced proper motion diagram of the LEHPMS stars closer than 25~pc. Squares: white dwarf candidates. Circles: halo subdwarf candidates. Triangles: M dwarf candidates. The solid lines indicate the limits we defined to isolate the white dwarfs and subdwarfs.}
   \label{fig4}
\end{figure}

Our nearby candidates listed in Table~1 all lie in the right part of the colour-reduced proper motion diagram where stars are most probably disc stars (Fig~\ref{fig4}). Thus their distance has been determined with the theoretical relation with the correct metallicity. 
But the use of a theoretical relation ($I-J$,$M_I$) at solar metallicity leads to an overestimate of the distances of the old population stars. According to \cite{baraffe1998} theoretical relations at different metallicities, the stars with the mean halo metallicity ($-1.8$ dex) are expected to be 3 to 4 magnitudes brighter than those with solar metallicity. 

Using the colour-reduced proper motion diagram, \cite{pokorny2003} found that about 20\% of the stars in the LEHPMS are halo subdwarfs. 
To isolate the halo subdwarfs using the reduced proper motion is even more difficult than the separation of white dwarfs as described in Sect.~\ref{xid}. We used both ($I-J$,$H_I$) and ($B_j-I$,$H_{B_j}$) diagrams to try to isolate halo subdwarfs.
We computed their distance using a theoretical colour-magnitude relation at metallicity $-1.8$ dex, interpolated from the \cite{lejeune1997} and \cite{baraffe1998} relations at different metallicities. Three stars that could be halo subdwarfs are found to be closer than 25~pc. Their characteristics are given at the end of Table~1 and Fig~\ref{fig4} shows their locus in the colour-reduced proper motion diagram. Most of them lie close to the limit we defined to isolate the subdwarfs. Our sample probably contains misidentified halo subdwarfs also close to the limit but on the right hand side.

\section{Conclusion}
We made a systematic search for nearby stars by cross-identifying the Liverpool-Edinburgh high proper motion survey with the DEep Near-Infrared Survey DENIS. The photometric distance is determined using DENIS photometry. 100 stars are found to be closer than the 25~pc limit of the CNS3. 10 are white dwarfs, 3 may be halo subdwarfs, the remainder are disc M dwarfs, mainly from spectral type M3 to M6. It is the first distance determination for 83 among them. 20 stars are found to be closer than 15~pc, of which for 10 stars there was no previously known distance: L 170-14A, DENIS J0146291-533931, DENIS J0235219-240038, LP 942-107, DENIS J0320518-635148, LTT 1732, DENIS J0428054-620929,  DENIS J2210200-701005, DENIS J2230096-534445, and NLTT 166-3. If it is a halo subdwarf, DENIS J2343155-241047 may also be closer than 15~pc.

Given the large uncertainties in the photometric distances (up to 45\% in some cases), follow-up observations of these candidates are needed. Low-resolution spectroscopy would assess the spectral type and provide more accurate determinations. This work should help in increasing the solar neighbourhood census. The nearest stars provide accurate data on fundamental parameters of stellar physics, such as luminosity, temperature, and mass. In particular, our understanding of low-mass stars relies upon the nearby stars, as they provide the only sample of intrinsically faint stars. They also are interesting targets for the future missions Terrestrial Planet Finder and DARWIN that will concentrate on very nearby stars to search for Earth-like exoplanets.

\begin{acknowledgements}
The authors thank Claire Ferrier for her participation to this work during her undergraduate studies. The use of Simbad and Vizier databases at CDS, as well as the ARICNS database was very helpful for this research.
The authors wish to thank the whole DENIS staff and all observers who collected the data. The DENIS project is supported by the SCIENCE and the Human Capital and Mobility plans of the European Commission under grants CT920791 and CT940627 in France, by the Institut National des Sciences de l'Univers, the Minist\`ere de l'\'education Nationale and the Centre National de la Recherche Scientifique (CNRS) in France, by the State of Baden-W\"urtemberg in Germany, by the DGICYT in Spain, by the Sterrewacht Leiden in Holland, by the Consiglio Nazionale delle Ricerche (CNR) in Italy, by the Fonds zur F\"orderung der wissenschaftlichen Forschung and Bundesministerium f\"ur Wissenschaft und Forschung in Austria, and by the ESO C \& EE grant A-04-046.
\end{acknowledgements}

\end{document}